**FIELDS, PARTICLES, NUCLEI**

# Gell-Mann–Low Function in the $\varphi^4$ Theory

## I. M. Suslov

*Kapitza Institute for Physical Problems, Russian Academy of Sciences, ul. Kosygina 2, Moscow, 117973 Russia*
*e-mail: suslov@kapitza.ras.ru*
Received November 1, 1999; in final form, February 2, 2000

An algorithm is proposed for the determination of the asymptotics of a sum of a perturbation series from the given values of its coefficients in the strong-coupling limit. When applied to the $\varphi^4$ theory, the algorithm yields the $\beta(g) \propto g^\alpha$ behavior with $\alpha \approx 1$ at large $g$ for the Gell-Mann–Low function. © 2000 MAIK "Nauka/Interperiodica".

Many problems in theoretical physics require advance in the strong-coupling region. The best known of them concern the dependence of the effective coupling constant $g$ on the distance scale $L$; the problems of electrodynamics at ultrashort distances and confinement are among them. The dependence of $g$ on $L$ in the renormalizable theories is determined by the equation

$$-dg/d\ln L = \beta(g) \qquad (1)$$

and generally requires information on the Gell-Mann–Low function $\beta(g)$ for arbitrary $g$ [1]. Over many years, the problem of reconstruction of the $\beta$-function seemed to be absolutely hopeless because the information on this function was provided solely by perturbation theory, which allowed the calculation of the first several terms of the expansion

$$\beta(g) = \sum_{N=0}^{\infty} \beta_N (-g)^N$$
$$= \beta_2 g^2 - \beta_3 g^3 + \ldots + \beta_N (-g)^N + \ldots, \qquad (2)$$
$$\beta_0 = \beta_1 = 0.$$

Lipatov [2] proposed a method allowing the calculation of the $\beta_N$ asymptotics at large $N$, which was found to be factorial for most problems:

$$\beta_N^{as} = ca^N \Gamma(N+b) \approx ca^N N^{b-1} N! \qquad (3)$$

Matching the Lipatov asymptotics (3) with the first $\beta_N$ coefficients provides information on all terms of the series and makes it possible to approximately reconstruct the $\beta$-function, but this requires a special procedure for the summation of divergent series [3]. Kazakov *et al.* [4] attempted to implement this procedure and arrived at the conclusion that the Gell-Mann–Low function in the $\varphi^4$ theory with the action functional

$$S\{\varphi\} = \int d^4x \left\{ \frac{1}{2}(\partial\varphi)^2 + \frac{16\pi^2}{4!} g\varphi^4 \right\} \qquad (4)$$

behaves at large $g$ as $0.9g^2$, which differs only in the coefficient from the one-loop result $1.5g^2$ valid at $g \longrightarrow 0$; similar behavior was obtained for $\beta(g)$ by Kubyshin [5]. If this result is valid,[1] then the $\varphi^4$ theory is self-contradictory. This conclusion seems to be strange from the viewpoint of solid-state applications: a reasonable model of a disordered system [7, 8], well-defined in the continuous limit, is mathematically reduced to the $\varphi^4$ model. Moreover, it was recently proved [9] that there are no renormalization singularities in the $\varphi^4$ theory; and this can be treated as evidence for the self-consistency of the theory.

This paper is aimed at revising the results obtained in [4, 5]. We start with the same premises as in [5], i.e., with the known first four coefficients of the $\beta$-function expansion [6, 10]

$$\beta(g) = \frac{3}{2}g^2 - \frac{17}{6}g^3 + \frac{154.14}{8}g^4 - \frac{2338}{16}g^5 + \ldots, \qquad (5)$$

and the Lipatov asymptotics with the first-order correction term calculated in [11]:

$$\beta_N = \frac{1.096}{16\pi^2} N^{7/2} N! \left\{ 1 - \frac{4.7}{N} + \ldots \right\}. \qquad (6)$$

The method is different from [4, 5] in that a direct relation between the $\beta(g)$ asymptotics and the expansion coefficients is used and the interpolation is carried out in an explicit form.

---

[1] The authors of [4] do not insist on their statement and emphasize that it has a tentative character (see also [6]).

**1.** Let us formulate the problem of reconstruction of the β-function asymptotics

$$\beta(g) = \beta_\infty g^\alpha, \quad g \longrightarrow \infty \quad (7)$$

from the coefficients $\beta_N$ of series (2) that grow according to factorial law (3) and are assumed to be numerically specified. As in the case of the introduction of critical indices in phase-transition theory, the slow (logarithmic) corrections to law (7) are considered to be beyond accuracy.

Treating the sum of series (2) in the Borel sense, we use a modified definition of the Borel transform $\tilde\beta(g)$:

$$\beta(g) = \int_0^\infty dx\, e^{-x} x^{b_0-1} B(gx), \quad B(g) = \sum_{N=0}^\infty B_N(-g)^N,$$

$$B_N = \frac{\beta_N}{\Gamma(N+b_0)}, \quad (8)$$

where $b_0$ is an arbitrary parameter that is conveniently used for optimizing the summation procedure [3]. As was assumed in [3] and proved recently in [9], the Borel transform is analytical in the complex $g$ plane with a cut from $-1/a$ to $-\infty$. To analytically continue $B(g)$ from the convergence circle $|g| < 1/a$ to arbitrary complex $g$ values, a conformal transformation $g = f(u)$ mapping the plane with the cut onto the unit circle is used; in this case, the reexpansion of $B(g)$ in $u$ powers

$$B(g) = \sum_{N=0}^\infty B_N(-g)^N \Big|_{g=f(u)}$$

$$\longrightarrow B(u) = \sum_{N=0}^\infty U_N u^N \quad (9)$$

gives a series convergent at arbitrary $g$ values. We restrict ourselves to the analytic continuation of $B(g)$ to the positive semiaxis [which is sufficient for the integration in Eq. (8)] and use a modified conformal transformation $g = (u/a)/(1-u)$ mapping the plane with cut $(-1/a, -\infty)$ onto the plane with cut $(1, \infty)$. This removes the $g = -1/a$ singularity to infinity, while the $g = \infty$ singularity becomes the nearest to the origin and determines the following asymptotics for the $U_N$ coefficients:

$$U_N = \frac{\beta_\infty}{a^\alpha \Gamma(\alpha)\Gamma(b_0+\alpha)} N^{\alpha-1}, \quad N \longrightarrow \infty. \quad (10)$$

This result can easily be obtained by representing the expansion coefficients as

$$U_N = \oint_C \frac{du}{2\pi i} \frac{B(u)}{u^{N+1}} \quad (11)$$

and deforming the contour $C$ enveloping the point $u=0$ in such a way that it passes around the cut with allowance made for the singularity $B(u) \sim (1-u)^{-\alpha}$ at the point $u=1$. The reexpansion of series (9) gives the following relation between $U_N$ and $B_N$:

$$U_0 = B_0,$$

$$U_N = \sum_{K=1}^N B_K \left(-\frac{1}{a}\right)^K C_{N-1}^{K-1} \quad (N \ge 1). \quad (12)$$

As a result, we arrive at the following simple algorithm: the coefficients $B_N$ are calculated from given $\beta_N$ [cf. Eq. (8)] and recalculated to $U_N$ according to Eq. (12); then the $U_N$ coefficients at large $N$ are fitted to the power law $U_N = U_\infty N^{\alpha-1}$, whose parameters determine $\beta_\infty$ and $\alpha$ according to Eq. (10). The $\beta_\infty$ value is convenient to calculate by treating $U_\infty$ as a function of $b_0$ and determining the slope of linear dependence $U_\infty \propto (b_0 + \alpha)$ at small $b_0 + \alpha$; this provides an independent estimate for the $\alpha$ index from the root of the $U_\infty(b_0)$ function.

**2.** The authors of the majority of works formulated the algorithm in such a way as to avoid mention of the coefficients $\beta_N$ for intermediate $N$ values. Such an approach is conceptually inconsistent, because a finite number of coefficients and their asymptotics can ensure the construction of a function with any prescribed behavior at infinity.[2] The problem can be reasonably formulated if *all* $\beta_N$ are approximately defined; in this case, the function $\beta(g)$ can be reconstructed within a certain accuracy. For this reason, the interpolation and estimation of its accuracy is the necessary step in solving the problem. Of course, this is possible only on the assumption that $\beta_N$ is a smooth function of $N$.

The interpolation is convenient to carry out for the reduced coefficient function

$$F_N = \frac{\beta_N}{\beta_N^{as}} = 1 + \frac{A_1}{N} + \frac{A_2}{N^2} + \ldots + \frac{A_K}{N^K} + \ldots, \quad (13)$$

which varies within the finite limits and has a regular expansion in $1/N$. Retaining in the series a finite number of terms and adjusting the coefficients $A_K$ to the known $F_N$ values, one obtains the desired interpolation formula. Its accuracy is

$$\delta F_N \sim \frac{A_{m+m_0+1}}{N^{m+m_0+1}} \frac{(N-L_0)(N-L_0-1)\ldots(N-L)}{L(L-1)\ldots L_0}, \quad (14)$$

if the interpolation is performed using $m$ known values $F_{L_0}, F_{L_0+1}, \ldots, F_L$ ($m = L - L_0 + 1$) for $m_0$ known coefficients $A_1, A_2, \ldots, A_{m_0}$. Estimate (14) is based on the fact that series (13) is asymptotic [12], so that the error of its approximation by truncation is of the order of the first omitted term, while the error of interpolating the $(m + m_0 + 1)$-order polynomial by the $(m + m_0)$-order polynomial can be calculated exactly. In the case under consideration, $L_0 = 2$, $L = 5$, and $m_0 = 1$, so that $\delta F_N$ is determined by the coefficient $A_6$, which can be estimated by factorial-law extrapolation of the found $A_1, \ldots, A_5$ values [12].

---

[2] A function of a factorial series has the same asymptotics for coefficients (3), but with different $c$ value [8]; the statement formulated in the text can easily be proved by choosing an appropriate linear combination of several functions.

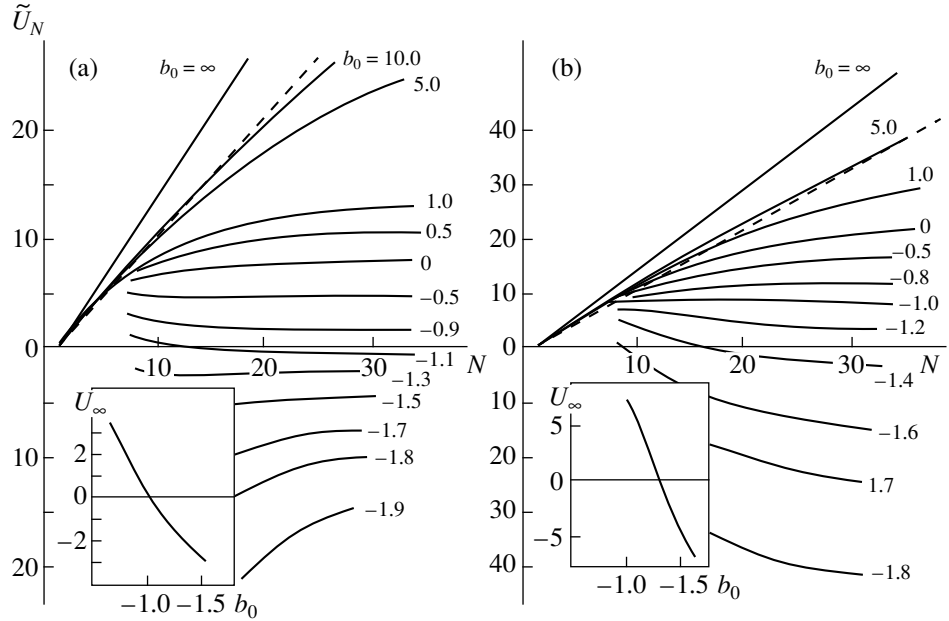

**Fig. 1.** Plots of $\tilde{U}_N = U_N \Gamma(b_0 + 2)$ vs. $N$ at different $b_0$ values for the parametrization of the asymptotics in the form (a) $\beta_N^{as} = ca^N N^{b-1} N!$ and (b) $\beta_N^{as} = ca^N \Gamma(N + b)$. The inserts show $U_\infty$ as a function of $b_0$.

The ambiguity of the interpolation procedure is manifested, in particular, in the possibility to differently parametrize the asymptotics [4, 13], e.g., as $ca^N \Gamma(N + b)$, $ca^N N^{b-1} N!$ etc. The asymptotics in the instanton calculations [2] has the form $\tilde{c}\tilde{a}^N N^{\tilde{b}} N^N$, which is very close to the Lipatov parametrization $ca^N N^{b-1} N!$ obtained from the former by applying the Stirling formula, whose accuracy is better than 10% even at $N = 1$. With this respect, the parametrization $ca^N N^{b-1} N!$ is "natural," whereas its representation in alternative functional forms requires additional assumptions [e.g., $N \gg b$ for $ca^N \Gamma(N + b)$].

Figure 1a shows the coefficients $\tilde{U}_N = U_N \Gamma(b_0 + 2)$ (normalized so that they have a finite limit at $b_0 \longrightarrow \infty$) for the natural parametrization $ca^N N^{b-1} N!$. At large $N$, these coefficients distinctly tend to the constant values (excepting the curves for $b_0 \gg 1$ and $b_0 \approx -2$, for which the large parameters retard the attainment of asymptotics), which corresponds to the value $\alpha = 1$. The $U_\infty$ vs. $b_0$ curve goes through zero at $b_0 = -1.03$ (see insert in Fig. 1a), which gives another estimate $\alpha = 1.03$ demonstrating excellent consistency of the results. Determining $\beta_\infty$ from the slope at $b_0 = -\alpha$, one obtains for the asymptotics of the $\beta$-function

$$\beta(g) \approx 8g \quad g \longrightarrow \infty. \quad (15)$$

A situation occurring at the alternative ways of interpolation is shown in Fig. 1b, where the parametrization $\beta_N^{as} = ca^N \Gamma(N + b)$ is used. There is also seen curve flattening, but it is not as distinct as in the preceding case. The processing of the curves under the assumption that $\alpha = 1$ yields the $b_0$ dependence of $U_\infty$ (see insert in Fig. 1b) going through zero at $b_0 = -1.3$, which corresponds to $\alpha = 1.3$. Hence, the results show a substantial inconsistency. Curve processing with the power-law dependence yields an $\alpha$ value slightly exceeding unity (different for different $b_0$), but in this case $U$ turns to zero at $b_0 = -0.8$ (at this value, the increase in the curves in Fig. 1b changes to a decrease), leading to the same inconsistency. Correspondingly, the result for the $\beta(g)$ asymptotics becomes less defined, $\beta(g) \approx 24g^\alpha$ and $\alpha = 0.8–1.3$.

Parametrizing the asymptotics as $\beta_N^{as} = ca^N N^{b-\tilde{b}-1} \Gamma(N + \tilde{b} + 1)$ and expanding $F_N$ in inverse powers of $(N - N_0)$, one obtains a two-parameter ($\tilde{b}$ and $N_0$) set of the interpolation formulas. The table presents the results for several such interpolations, for which the distinctions in the interpolation curves approximately fit the error range estimated by Eq. (14) for the natural interpolation. With allowance made for the uncertain-ties, the $\alpha$ index is virtually independent of the particular interpolation; systematic deviations occur only for some "extreme" cases for which the interpolation curve is partially beyond the error range. The interpolation with $\tilde{b} = 0$ and $N_0 = 0$, which we treated as "natural" from computational considerations, stands out as the most self-consistent algorithm. Therefore, the corresponding result (15) should be considered as the most reliable. For a fixed interpolation, its error is less than 0.05 for the $\alpha$ index and 10% for $\beta_\infty$, which is an optimistic estimate for the accuracy. The error caused by the interpolation ambiguity is seen from the table: it can be as great as several tenths for the $\alpha$ index, whereas the $\beta_\infty$ value can differ from Eq. (15) by a factor of 2–3.

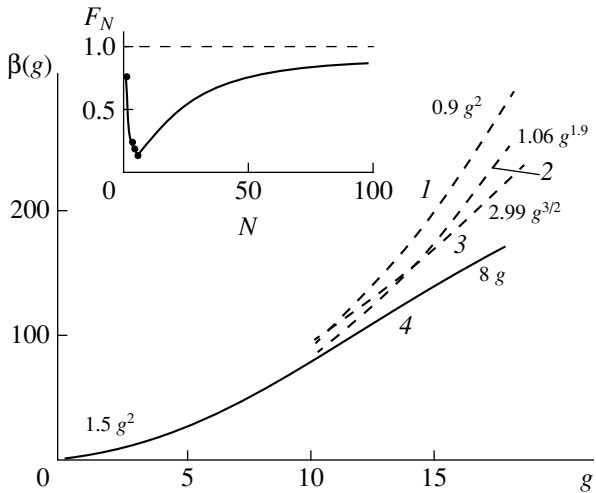

**Fig. 2.** Qualitative behavior of the Gell-Mann–Low function in the $\varphi^4$ theory. Curves *1*, *2*, *3*, and *4* are obtained in [4], [5], [14], and this work, respectively. The insert shows the coefficient function $F_N = \beta_N/\beta_N^{as}$ with the $\beta_N^{as} = ca^N\Gamma(N + b)$ parametrization convenient for an analysis of Eq. (12); the dependence on the interpolation procedure is immaterial on this scale.

**3.** Let us consider the behavior of the $\beta$-function at finite $g$ values. For $N < 10$, Fig. 1 demonstrates a linear portion $\tilde{U}_N \approx 1.1(N - 1)$ (dashed line) corresponding to a $\beta(g) \approx 1.1g^2$ dependence close to the results obtained in [4, 5]. This portion is insensitive to changes in $b_0$ and to the interpolation procedure and can pretend to the role of the true asymptotics, provided that the results for $N > 10$ are treated as being due to the interpolation errors. But such is actually not the case, because the stability of this portion is caused by the presence of a characteristic dip in the reduced coefficient function $F_N$ at $N \lesssim 10$ (see insert in Fig. 2). If one models this dip by setting $F_3 = F_4 = \ldots = F_{10} = 0$, then the result $\tilde{U}_N \approx 1.5(N - 1)$ determined by the first nonvanishing coefficient $F_2$ (see curve for $b_0 = \infty$) is obtained at $N \leq 10$ for all $b_0$; this is close to the actual situation.[3] Such modeling of the dip demonstrates that the one-loop law $1.5g^2$ for the $\beta$-function extends up to $g \sim 10$. More precisely (see footnote 3), the result valid in the interval $1 \lesssim g \lesssim 10$ is given by the function derived in [4, 5] and yielding the value $\beta(g) \approx 90$ for $g = 10$ (see Fig. 2), in accordance with [14]. Asymptotics Eq. (15) matches well with the indicated value, providing indirect support to the optimistic estimation for the accuracy.

Although the available information allows only a rough estimation for the Gell-Mann–Low function, one can state with assurance that it is nonzero at finite $g$ values and its behavior at $g \longrightarrow \infty$ is compatible with the assumption that the $\varphi^4$ theory is self-consistent. The substitution of Eq. (15) in Eq. (1) yields the $g(L) \propto L^{-\gamma}$ dependence with $\gamma \approx 8$ at small $L$, which is slightly modified if the $\alpha$ index is other than unity or if logarithmic branching is present.

The results obtained allow an understanding of why the numerical simulations on a lattice indicate that the $\varphi^4$ theory is "trivial" (see [15] and references therein): because of the absence of zeros of the $\beta$-function, the $g(L)$ interaction always decreases with distance; and, owing to the extended one-loop law, the behavior is indistinguishable from the trivial in a wide range of parameters [at $g \lesssim 300$, for the most popular charge definition when the term with interaction in Eq. (4) has the form $g\varphi^4/4$].

This work was supported by the INTAS (grant no. 96-0580) and the Russian Foundation for Basic Research (project no. 00-02-17129).

**Table**

| Interpolation with $N_0 = 0$ | $\beta_\infty$ | $\alpha$ | Interpolation with $\tilde{b} = 0$ | $\beta_\infty$ | $\alpha$ |
|---|---|---|---|---|---|
| $\tilde{b} = 3.5$ | 24 | 0.8–1.3 | $N_0 = 0.5$ | 5.4 | 1.1–1.6 |
| $\tilde{b} = 1.5$ | 14 | 1.0–1.1 | $N_0 = 0.3$ | 16 | 0.8–1.2 |
| $\tilde{b} = 0$ | 8.1 | 1.0 | | | |
| $\tilde{b} = -1.5$ | 10 | 1.0–1.1 | $N_0 = -0.3$ | 4.3 | 0.9–1.0 |
| $\tilde{b} = -2.5$ | 2.7 | 1.5–1.7 | $N_0 = -0.5$ | 2.3 | 0.9–1.0 |

REFERENCES

1. N. N. Bogoliubov and D. V. Shirkov, *Introduction to the Theory of Quantized Fields*, 3rd ed. (Nauka, Moscow, 1984, 4th ed.; Wiley, New York, 1980, 3rd ed.).

---

[3] If one sets $F_3 = F_4 = \ldots = F_{10} = \epsilon$, then Eq. (12) yields a linear dependence with the slope $1.5(1 - \epsilon/F_2)$ for $b_0 = b - p$ with integer $p$ in the interval $p + 2 \leq N \leq 10$. For $\epsilon = 0.2$ (see Fig. 2); the correct slope 1.1 and a weak dependence on $b_0$ occur in the interval $-1 < b_0 < 10$.

2. L. N. Lipatov, Zh. Éksp. Teor. Fiz. **72**, 411 (1977) [Sov. Phys. JETP **45**, 216 (1977)].
3. J. C. Le Guillou and J. Zinn-Justin, Phys. Rev. Lett. **39**, 95 (1977); Phys. Rev. B **21**, 3976 (1980).
4. D. I. Kazakov, O. V. Tarasov, and D. V. Shirkov, Teor. Mat. Fiz. **38**, 15 (1979).
5. Yu. A. Kubyshin, Teor. Mat. Fiz. **58**, 137 (1984).
6. A. A. Vladimirov and D. V. Shirkov, Usp. Fiz. Nauk **129**, 407 (1979) [Sov. Phys. Usp. **22**, 860 (1979)].
7. M. V. Sadovskiĭ, Usp. Fiz. Nauk **133**, 223 (1981) [Sov. Phys. Usp. **24**, 96 (1981)].
8. I. M. Suslov, Usp. Fiz. Nauk **168**, 503 (1998) [Phys. Usp. **41**, 441 (1998)].
9. I. M. Suslov, Zh. Éksp. Teor. Fiz. **116**, 369 (1999) [JETP **89**, 197 (1999)].
10. F. M. Dittes, Yu. A. Kubyshin, and O. V. Tarasov, Teor. Mat. Fiz. **37**, 66 (1978).
11. Yu. A. Kubyshin, Teor. Mat. Fiz. **57**, 363 (1983).
12. I. M. Suslov, Zh. Éksp. Teor. Fiz. **117** (2000) (in press).
13. V. S. Popov, V. L. Eletskiĭ, and A. V. Turbiner, Zh. Éksp. Teor. Fiz. **74**, 445 (1978) [Sov. Phys. JETP **47**, 232 (1978)].
14. A. N. Sissakian, I. L. Solovtsov, and O. P. Solovtsova, Phys. Lett. B **321**, 381 (1994).
15. A. Agodi, G. Andronico, P. Cea, *et al.*, Mod. Phys. Lett. A **12**, 1011 (1997); hep-ph/9702407.